\documentclass[journal=jpccck,manuscript=article]{achemso}

\usepackage{amsmath}

\author{David F. Macias-Pinilla}
\affiliation[QFA]
{Departament de Qu\'imica F\'isica i Anal\'itica, Universitat Jaume I, Av. Sos Baynat, s/n, 12071 Castell\'o, Spain}
\alsoaffiliation[INAM]
{Institute of Advanced Materials (INAM), 
 Universitat Jaume I, Av. Sos Baynat, s/n, 12071 Castell\'o, Spain}

\author{Josep Planelles}
\affiliation[QFA]
{Departament de Qu\'imica F\'isica i Anal\'itica, Universitat Jaume I, Av. Sos Baynat, s/n, 12071 Castell\'o, Spain}

\author{Iv\'an Mora-Ser\'o}
\affiliation[INAM]
{Institute of Advanced Materials (INAM), 
 Universitat Jaume I, Av. Sos Baynat, s/n, 12071 Castell\'o, Spain}

\author{Juan I. Climente}
\affiliation[QFA]
{Departament de Qu\'imica F\'isica i Anal\'itica, Universitat Jaume I, Av. Sos Baynat, s/n, 12071 Castell\'o, Spain}
\email{climente@uji.es}

\title{Comparison between trion and exciton electronic properties in CdSe and PbS nanoplatelets}

\begin{document}

\begin{tocentry}
\begin{center}
\includegraphics[width=8.25cm]{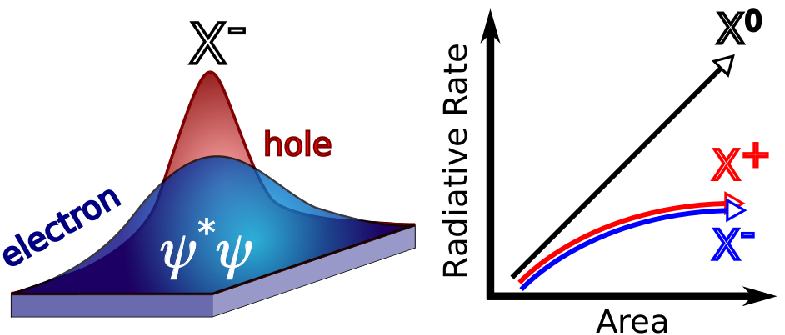} 
\end{center}
\end{tocentry}

\begin{abstract}
The optoelectronic properties of metal chalcogenide colloidal nanoplatelets are often interpreted 
in terms of excitonic states. However, recent spectroscopic experiments evidence the presence 
of trion states, enabled by the slow Auger recombination in these structures.
We analyze how the presence of an additional charge in trions modifies the emission energy
 and oscillator strength as compared to neutral excitons.
These properties are very sensitive to dielectric confinement and electronic correlations,
which we describe accurately using image-charge and variational Quantum Monte Carlo methods
in effective mass Hamiltonians. 
We observe that the giant oscillator strength of neutral excitons is largely suppressed
in trions. 
Both negative and positive trions are redshifted with respect to the exciton,
 and their emission energy increases with increasing dielectric mismatch between the 
platelet and its surroundings, which is a consequence of the self-energy potential.
Our results are consistent with experiments in the literature, and assess on the 
validity of previous theoretical approximations.
\end{abstract}


\section{\label{sec:level1} Introduction}

The outstanding optical features of semiconductor colloidal nanoplatelets (NPLs)
makes them a broad target of study and application in devices.\cite{ZhoyingAFM,diroll2020,DuttaPCC,nasilowski2016} 
As compared to colloidal quantum dots, the quasi-2D nature of NPLs conveys
several distinct properties of interest, including precisely controllable 
thickness --which enables narrow emission linewidths--, 
large absorption cross-sections and bright luminescence resulting from 
the giant oscillator strength effect.\cite{diroll2020,DuttaPCC,nasilowski2016}
Several efforts have been carried out in synthesis methods to control the morphology,
dimensions and composition, as well as in characterization of the subsequent 
optoelectronic properties.\cite{nasilowski2016,zhang2020heterostructures,pun2021understanding,min2019} 
In parallel, theoretical models have been developed to understand the underlying electronic structure.
In most cases, such models focus on neutral excitons,\cite{ithurria2011colloidal,achtstein2012electronic,Benchamekh2014,
yang2015electronic,bose2016effect,rajadell2017excitons,richter2017nanoplatelets,polovitsyn2017synthesis,
chu2018electronic,achtstein2018impact,specht2019size,llusar2019strain,steinmetz2020emission,macias2021morphology,Greenwood2021}
which is typically the dominant emitting species in colloidal nanocrystals. 


There is however increasing evidence that the emission spectrum of core-only and core/shell CdSe NPLs 
at low temperatures has a significant contribution from charged excitons (trions).\cite{Lorenzon2015,yu2019effect,Shornikova2020,antolinez2019observation,antolinez2020trion,peng2020bright,ayari2020tuning}
The formation of radiative trions is enabled by the higher dimensionality of NPLs as compared to nanocrystals,
which slows down Auger relaxation processes.\cite{kunneman2013,baghani2015,li2017,philbin2020}
Trions are of high technological interest for optoelectronic applications.
The spin-zero configuration in one of the bands suppresses the electron-hole exchange interaction,
and hence the optically dark ground state observed in the fine structure of neutral excitons,
which is convenient for efficient luminescence.\cite{Shornikova2020}
They also enable two-color emission through shake-up processes\cite{lobl2020radiative} --which are particularly strong in colloidal
NPLs\cite{antolinez2019observation,llusar2020nature}--, provide a photoresponsive system for chemical sensing through surface 
adsorption\cite{Lorenzon2015} and form a natural two-level quantum system for quantum information applications.\cite{peng2020bright}

Understanding how the electronic structure of trions differs from that of excitons is a prerequesite for rational design
of applications. 
In a recent work, Peng \emph{et al.} analyzed the emission spectra of exciton and trion states in 
CdSe NPLs with $4.5$ monolayers core thickness surrounded by thin CdS/ZnS shells.\cite{peng2020bright} 
By comparing time-resolved photoluminescence data with tight-binding simulations, 
 which accounted for electronic correlations using a configuration interaction (CI) approach, 
they assigned the observed spectral peaks to neutral excitons and negative trions. 
The latter showed binding energies of about $10.5$ meV, and similar oscillator strength to that 
of the exciton band edge transition.
 In parallel, Ayari \emph{et al.} investigated the negative trion emission spectra in CdSe core-only NPLs with 
the same thickness, but varying lateral dimensions.\cite{ayari2020tuning} 
They observed stronger trion binding energies, widely modulated by the lateral confinement (18-36 meV).
By comparing time-integrated photoluminescence data with rate equation models
and effective mass-CI calculations of the electronic structure,
 they inferred that the oscillator strength of trions is one order of magnitude weaker than that of the excitons.  
 The seemingly contradicting conclusions in the two studies call for independent assessment. 

To better understand how the photophyics of trions in NPLs differs from that of excitons, one must account 
for the fact that electronic correlations are particularly challenging to model in these structures.
Unlike quantum dots, NPLs lie in a Coulomb dominated, intermediate confinement regime,\cite{rajadell2017excitons,richter2017nanoplatelets}
where the strong electron-hole attraction --arising from dielectric confinement and the quasi-2D geometry-- 
leads to small excitonic radii.\cite{brumberg2019determination}
Such a short-range interaction makes CI calculations based on envelope wave functions
miss correlation energy.\cite{planelles2017simple,Rontani_2017}
By contrast, the electron-electron (or hole-hole) repulsions in trions, 
which are also enhanced by the quasi-2D geometry and the dielectric confinement,
are longer-ranged interactions and hence better captured by CI models. 
Conclusions about the balance between attractive and repulsive
forces in these systems must be revisited using methods which 
account for both of them on equal footing.

In this work, we calculate the electronic structure of excitons and trions in colloidal NPLs 
using effective mass Hamiltonians which include electronic correlations though a variational
Quantum Monte Carlo (VQMC) description. 
This method has been recently shown to outperform CI approaches in these systems.\cite{planelles2021simple}
We choose CdSe and PbS NPLs, which are systems of current interest for direct applications including 
solar cells\cite{carey2015colloidal,sanchez2020preferred}, photodetectors operating in the near infrared \cite{DeIacovo2016}, light-emitting diodes \cite{sanchez2014all,fan2015colloidal,van2020luminescence} and lasers \cite{guzelturk2014amplified}. We provide an overview on the differences between neutral excitons and trions,
and examine some of the approximations used in earlier models, such as the description of dielectric 
confinement with Rytova-Keldysh (RK) potential and neglecting self-energy corrections.\cite{ayari2020tuning}
%
We shall see that of both positive and negative trions are bound for typical NPL dimensions,
with binding energies which (for CdSe) exceed room temperature thermal energy.
The repulsions in trions lead to a moderate increase of the excitonic Bohr radius and 
induce a significant suppression of the oscillator strength as compared to neutral excitons.
Similar to excitons, dielectric confinement leads to an overall increase of the trion
emission energy.

\section{Methods}
We calculate exciton and trion ground state energies and wave functions 
closely following Ref.~\citenum{planelles2021simple}.
Briefly, we solve effective mass Hamiltonians of the form:
\begin{equation} \label{eq:H}
	H =  
	\sum_{i=1}^{N_e} \hat{h}_e 
	+ \sum_{i=1}^{N_h} \hat{h}_h 
	+ \sum_{i=1}^{N_e} \sum_{j=1}^{N_h} V_c(\mathbf{r}_{i},\mathbf{r}_{j}) 
	+ 1/2 \sum_{i=1}^{N_e} \sum_{j \neq i}^{N_e} V_c(\mathbf{r}_{i},\mathbf{r}_{j}) 
	+ 1/2 \sum_{i=1}^{N_h} \sum_{j \neq i}^{N_h} V_c(\mathbf{r}_{i},\mathbf{r}_{j})  
	+E_{gap}.
\end{equation}
\noindent where $N_e$ and $N_h$ is the number of electrons and holes, 
while $\hat{h}_e$ and $\hat{h}_h$ are the corresponding single-particle Hamiltonians:
\begin{equation}
\hat{h}_i=\Big(\frac{\mathbf{\hat{p}}_{\perp}^2}{2m_{\perp,i}}+\frac{\hat{p}_{z}^2}{2m_{z,i}}+ V \Big).
\end{equation}
Here, $m_{\perp}$ and $\mathbf{\hat{p}_{\perp}}$ ($m_{z,i}$  and $\hat{p}_{z}^2$) are the in-plane (out-of-plane) 
effective mass and momentum operators, respectively. $V=V^{conf}+\Sigma$, is the single particle potential, 
with $V^{conf}$ the spatial confinament potential --zero inside the cuboidal NPL, infinite outside it--,
and $\Sigma$ the self-energy potential arising from dielectric confinement.
$V_c(\mathbf{r}_{i},\mathbf{r}_{j})$ are the Coulomb interaction terms and $E_{gap}$ the bulk band gap.

Dielectric confinement is important in colloidal NPLs, owing to the strong mismatch between the inorganic NPL
and the organic surroundings.
Because the organic environment screens electric fields more weakly than the semiconductor lattice, 
it gives rise to enhanced Coulomb interactions.\cite{achtstein2012electronic,rajadell2017excitons,Benchamekh2014}
We model this effect using image-charge (IC) expressions developed for finite-width quantum wells. 
Self-energy corrections ($\Sigma$) are included within the same formalism to account for the changes 
that individual carriers produce on their own dielectric environment.\cite{kumagai1989excitonic}
In some instances, for the sake of comparison, the Coulomb enhancement arising from dielectric confinement 
is modeled using a RK potential instead,
which is an approximation of the Coulomb potential for the limit case of a quantum well with zero 
thickness and static dielectric screening.\cite{cudazzo2011dielectric}

To solve Hamiltonian (\ref{eq:H}), a variational trial wave function is defined for excitons:
\begin{equation} \label{ewf}
	\Psi_{X}(\mathbf{r_e},\mathbf{r_h},\sigma_e,\sigma_h)= N_{X}\,\Phi_e(\mathbf{r_e})\Phi_h(\mathbf{r_h})J(r_{eh}), 
\end{equation}
\noindent where $N_X$ is the normalization constant, 
 $\Phi_e$ and $\Phi_h$ are the envelope functions of non-interacting electron and hole,
 and $J(r_{eh})=e^{- r_{eh} / r_B^{X} }$ is a Slater correlation factor, 
 with $r_{eh}$ the electron-hole separation and $r_B^{X}$ the effective exciton Bohr radius, 
 which is the variational parameter to optimize.
 For negative trions, the ground state involves double occupancy of the lowest electron orbital 
 (associated to singlet spin) plus one hole. Then, a Slater-Jastrow trial wave function is used: 
\begin{equation} \label{twf}
 \Psi_{X^-}(\mathbf{r_{e_1}},\mathbf{r_{e_2}},\mathbf{r_{h_1}},\sigma_{e_1},\sigma_{e_2},\sigma_{h_1}) = 
	N_{X^-}\,\Phi_{e}(\mathbf{r_{e_1}})\Phi_{e}(\mathbf{r_{e_2}})\Phi_{h}(\mathbf{r_{h_1}}) J(r_1,r_2,r_{12}) .
\end{equation}
\noindent Here, $N_{X^-}$ the normalization factor and 
%
$ J(r_1,r_2,r_{12})=e^{-r_1/r_B^{X^-}}\ e^{-r_{2}/r_B^{X^-}}\ e^{\frac{ b r_{12}}{1+ a r_{12}}}$
the Jastrow factor,
with $r_{1}=|\mathbf{r_{e_1}}-\mathbf{r_{h_{1}}}|$, 
$r_{2}=|\mathbf{r_{e_2}}-\mathbf{r_{h_1}}|$ and 
$r_{12}=|\mathbf{r_{e_1}}-\mathbf{r_{e_2}}|$. 
In this case, $r_B^{X^-}$, $b$ and $a$ are the variational parameters to optimize.
The first one is related to attractions and the others to repulsions.
Analogous expressions are used for positive trions.

Variational optimization is carried out using a VQMC algorithm, 
with the freely available codes we developed in Ref.~\citenum{planelles2021simple}. 
Iterative Newton-Rhapson convergence is compared with sequential optimization
of the variational parameters to avoid local minima.
The optical recombination probability for the resulting wave functions can be 
estimated within the dipole approximation.\cite{Pawel_book} 
For a neutral exciton, the transition from $|\Psi_X\rangle$ to a vacuum state $|0\rangle$
is proportional to the electron-hole overlap:\cite{rajadell2017excitons}
\begin{equation}
	\tau^{-1}_{X \rightarrow 0} \propto 1/2 \, |\langle 0 | \delta_{\mathbf{r_e},\mathbf{r_h}} | \Psi_X \rangle|^2 =
	\frac{1}{2} \, \left(\frac{N_X}{N_e\,N_h}\right)^2.
\end{equation}
\noindent where $N_i$ is the normalization constant of a single-particle orbital $\Phi_i$,
$N_e=N_h=\sqrt{8/(L_x L_y L_z)}$ in our particle-in-box functions,
with $L_x$, $L_y$ and $L_z$ the NPL dimensions. The $1/2$ factor accounts for
spin selection rules forbidding optical transitions in dark excitons.
For a negative trion, the transition from $|\Psi_X^-\rangle$ to 
an electron state $|\Psi_e\rangle = N_e \Phi_e(\mathbf{r_{e_1}})$
 is proportional to:
\begin{equation}
	\tau^{-1}_{X^- \rightarrow e} \propto \left|\langle \Psi_e | \delta_{\mathbf{r_{e_2}},\mathbf{r_{h_1}}} | \Psi_{X^-} \rangle \right|^2= 
	(N_e \, N_{X^-} \, \mathcal{I} )^2,
\end{equation}
\noindent with $\mathcal{I}=\int \Phi_e(\mathbf{r_{e_1}})^2 \, \Phi_e(\mathbf{r_{e_2}}) \, \Phi_{h}(\mathbf{r_{e_2}})\,
       e^{-r_{12}/r_B^{X^-}} \, e^{\frac{ b r_{12}}{1+ a r_{12}}} \, d^3r_{e_1} \, d^3r_{e_2}$.

\section{Results}

We investigate excitons and trions in CdSe and PbS NPLs with cubic crystal lattice.
Material parameters are summarized in Table \ref{T1}.
For CdSe, we consider NPLs with $4.5$ monolayer thickness ($L_z=1.4$ nm), 
as in the samples measured in Refs.~\citenum{peng2020bright} and \citenum{ayari2020tuning}. 
For PbS, we choose a typical thickness of $7-8$ monolayers ($L_z=2.0$ nm).\cite{antu2018bright}
Unless otherwise stated, for both CdSe and PbS NPLs we consider an environmental  
dielectric constant due to the organic ligands of $\varepsilon_{out}=2$.\cite{ludde1959dielektrische,achtstein2012electronic}. 

\begin{table}[!h]
\begin{tabular}{cllcc}
{\bf Parameter} & \hspace{0.2 cm}\begin{tabular}[c]{@{}c@{}}{\bf CdSe}\\\end{tabular} & \hspace{0.2 cm} \begin{tabular}[c]{@{}c@{}}{\bf PbS}\end{tabular} & \hspace{0.2 cm}\begin{tabular}[c]{@{}c@{}}{\bf Units }\\  \end{tabular}
                                                                                                                                    \\ \hline
$\varepsilon_{NPL}$  & \begin{tabular}[c]{@{}l@{}} 10 \cite{adachi2004handbook} \end{tabular} & 17 \cite{gupta1983analysis} & \begin{tabular}[c]{@{}l@{}} $\varepsilon_0$ \end{tabular}   \\ \hline
$m_{\perp,e}$  & \begin{tabular}[c]{@{}l@{}} 0.22 \cite{Benchamekh2014} \end{tabular}  & 0.27 \cite{macias2021morphology} & \begin{tabular}[c]{@{}l@{}}$m_0$ \end{tabular}   \\ \hline
 $m_{\perp,h}$ & \begin{tabular}[c]{@{}l@{}} 0.41 \cite{Benchamekh2014} \end{tabular}  & 0.19 \cite{macias2021morphology} & \begin{tabular}[c]{@{}l@{}} $m_0$\end{tabular}   \\ \hline
 $m_{z,e}$   & \begin{tabular}[c]{@{}l@{}} 0.4 \cite{rajadell2017excitons}  \end{tabular} & 0.29 \cite{macias2021morphology} & \begin{tabular}[c]{@{}l@{}} $m_0$\end{tabular}   \\ \hline
  $m_{z,h}$   & \begin{tabular}[c]{@{}l@{}} 0.9 \cite{rajadell2017excitons}  \end{tabular} & 0.25 \cite{macias2021morphology} & \begin{tabular}[c]{@{}l@{}} $m_0$\end{tabular}  \\ \hline
$E_{gap}$   & \begin{tabular}[c]{@{}l@{}} 1.76 \cite{adachi2004handbook}  \end{tabular} & 0.41 \cite{kang1997electronic} & \begin{tabular}[c]{@{}l@{}} eV\end{tabular}  \\ \hline
 $a_0$    & \begin{tabular}[c]{@{}l@{}} 6.08 \cite{adachi2004handbook}  \end{tabular} & 5.935 \cite{hoda1975phase} & \begin{tabular}[c]{@{}l@{}} \r{A} \end{tabular}  \\ \hline
\end{tabular}
\caption{Material parameters used in the calculations. $\varepsilon_0$ is the vacuum permittivity, $m_0$ is the free electron mass and $a_0$ stands for the lattice constant. The references indicate the data source. }
\label{T1}
\end{table}

\subsection{Binding Energy}

\begin{figure*}[h!]
\includegraphics{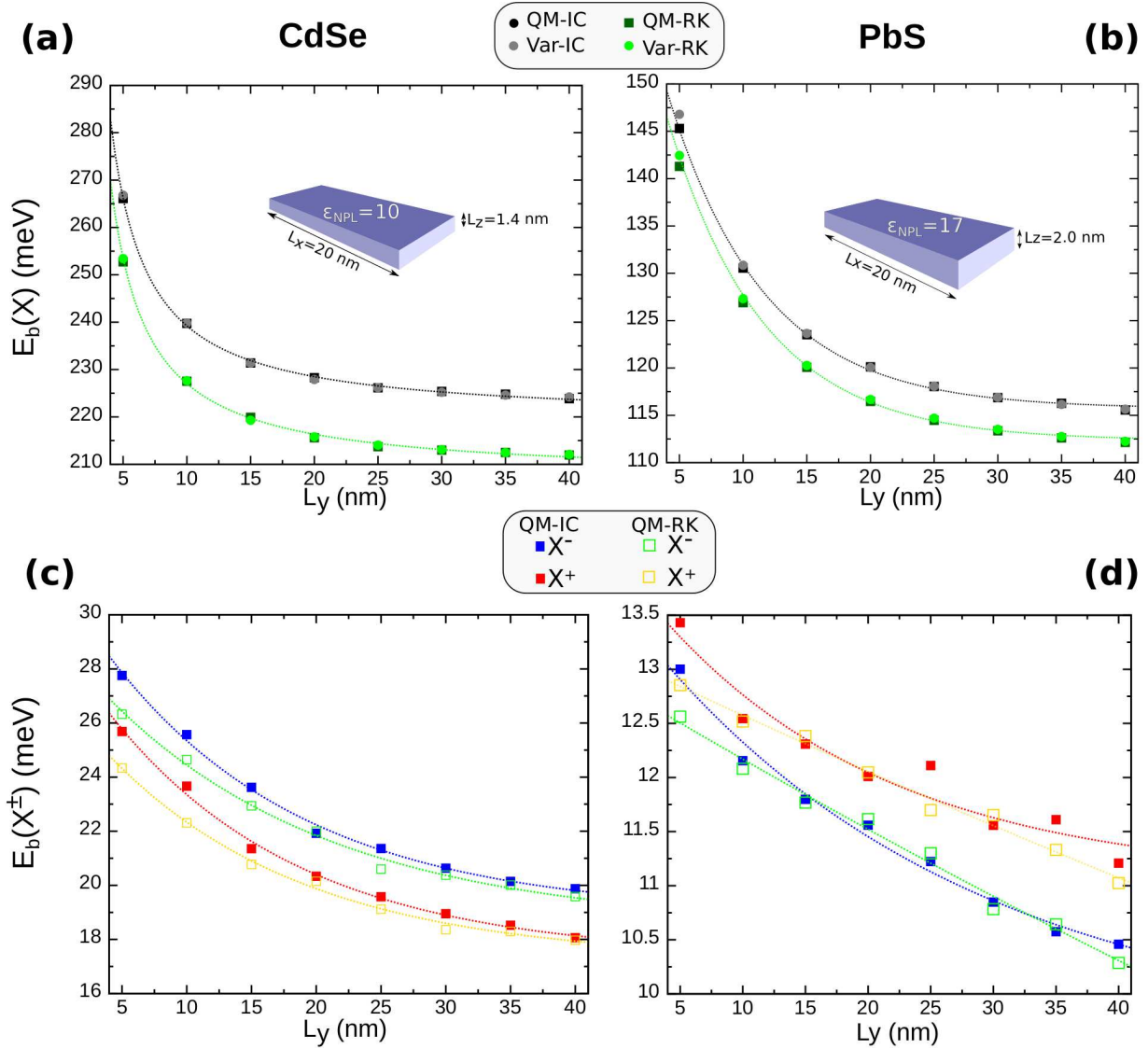}
\caption{\label{fig:1} 
Binding energies as a function of lateral NPL confinement for: 
(a) CdSe exciton, (b) PbS exciton, (c) CdSe trions and (d) PbS trions.
Squares are used for calculations using VQMC and circles for those using 
a semi-analytic variational calculation. 
IC and RK potentials describing dielectric confinement are also compared.
Lines are guides to the eye.
Fluctuations around them are due to the statistical nature of VQMC calculations.
Insets: schematics of the NPLs under study.}
\label{fig1}
\end{figure*}

We start by studying the effect of lateral confinement on the binding energy of excitons and trions.
For excitons, the binding energy is calculated as $E_b(X)=E_e+E_h-E(X)$, 
with $E(X)$ the the total exciton energy and $E_{e/h}$ that of an independent electron/hole. 
For the trion, it is defined as $E_b(X^{\pm})=E(X)+E_{h/e}-E(X^{\pm})$, 
where $E(X^{\pm})$ is the total energy of the (positive or negative) trion. 
We take a NPL with one fixed lateral side ($L_x=20$ nm) and vary the other side ($L_y$).
The results are represented as blue squares in Figure \ref{fig1}. 
In the case of excitons, we also plot the binding energies
obtained with a semi-analytic variational method\cite{rajadell2017excitons}, blue circles in Fig.~\ref{fig1}(a) and (b). 
The excellent agreement between the two methods validates the performance of the VQMC approach, 
which has the advantage of being computationally efficient for trions as well.

Figs.~\ref{fig1}(a) and (b) show that the exciton binding energies in NPLs are large, 
which is a result of the quasi-2D geometry and the strong dielectric confinement.
For CdSe NPLs, the values (225-265 meV) are consistent with previous theoretical estimates\cite{achtstein2012electronic,rajadell2017excitons,Benchamekh2014,ayari2020tuning},
 and experimental measurements.\cite{zelewski2019exciton} 
In PbS NPLs, the binding energies (115-145 meV) exceed those calculated by Yang and Wise
for quantum wells ($\sim 80$ meV).\cite{yang2015electronic} This is a consequence 
of the smaller thickness in our NPLs ($L_z=2$ nm vs. $L_z=3$ nm in their work),
and suggests that recent progress in the synthesis of thin PbS NPLs\cite{antu2018bright,khan2017near,akkerman2019} 
should facilitate the observation of robust excitonic effects.
 Also, our results evidence that PbS NPLs can benefit from a significant modulation of 
the binding energy with lateral confinement, starting at large
lateral dimensions well above the PbS excitonic Bohr radius ($L_y<30$ nm).

Figs.\ref{fig1}(c) and (d) show the calculated trion binding energies for
negative (blue squares) and positive (red squares) trions.
The modulation of $E_b(X^\pm)$ by lateral confinement is evident. 
In the case of CdSe, it enables binding energies greater than thermal energy 
at room temperature ($24$ meV) when $L_y<15$ nm, which is consistent with Ref.~\citenum{ayari2020tuning} experiments.
Even in the case of PbS, which has a large dielectric constant, the trion binding energies are
one order of magnitude greater than those of epitaxial quantum wells.\cite{esser2000photoluminescence}
For both materials and all dimensions under study, trions are bound ($E_b(X^\pm)>0$),
which should be seen in experiments as a redshifted emission when compared to excitons.
This is a consequence of the strong Coulomb correlation, and is in
contrast to small CdSe nanocrystals.\cite{peng2020bright,califano2007lifetime}

Previous theoretical simulations of trions in NPLs using effective mass formalism\cite{ayari2020tuning}
introduced dielectric effects in Coulomb interactions through a RK potential.\cite{ayari2021correction} 
Although computationally this is a very efficient approximation, the potential was originally
devised for strictly two-dimensional structures (graphene, transition metal dichalcogenides)\cite{cudazzo2011dielectric}
and its performance in colloidal NPLs, whose thickness can reach few nm, has not been tested.
In Fig.~\ref{fig1} we compare the binding energies calculated with RK and IC potentials.
For excitons, RK simultations (green squares in Fig.~\ref{fig1}(a,b)) underestimate the binding
energy by $5-10\%$ (cf. blue squares). For trions, however, RK and IC energies match within 1 meV 
difference. Since effective mass models are meant to provide qualitative and semi-quantitative
estimates, these results support the use of the RK approximation to calculate binding energies
in colloidal NPLs with typical (few monolayer) thicknesses.
Nonetheless, for estimates of the emission energies, the RK potential must be complemented
with self-energy corrections, as we show in the next section.

\subsection{Modulation of dielectric confinement}

\begin{figure*}[h!]
\includegraphics{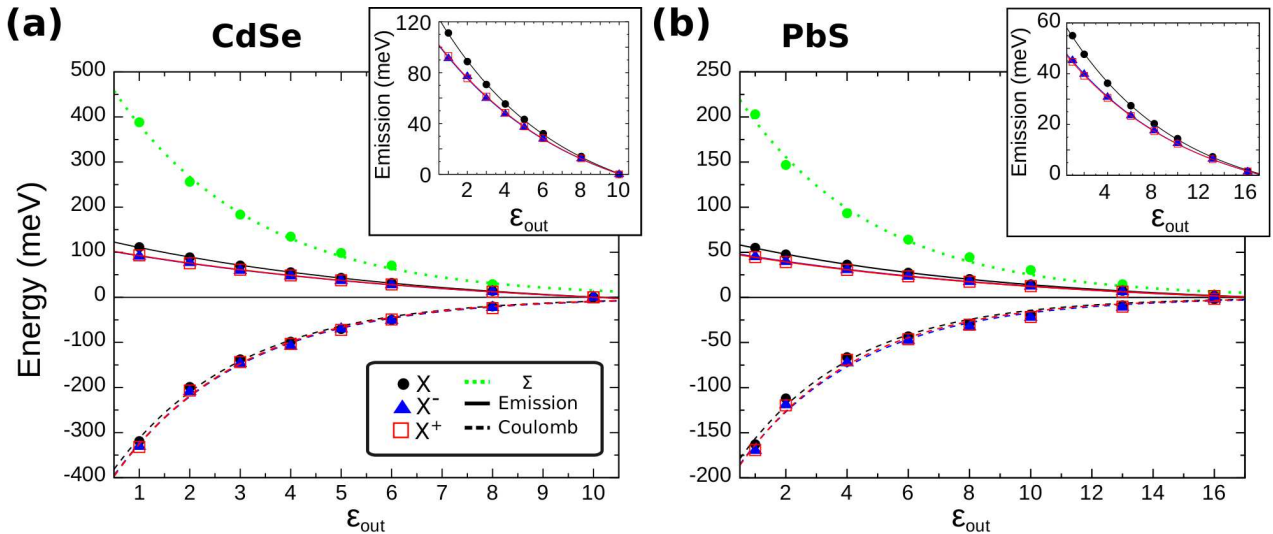}
\caption{\label{fig2} Shift of emission energy (solid line), net Coulomb energy (dashed line) and self-energy (dotted line)
as a function of the dielectric constant in the environment in (a) CdSe NPL and (b) PbS NPL. 
All energies are referred to the $\varepsilon_{out}=\varepsilon_{NPL}$ case.
Different symbols are used for excitons and trions. 
The NPL has $20\times10$ nm$^2$ area. 
The insets zoom in on the emission blueshift for excitons and trions, which results from 
$\Sigma$ exceeding the net Coulomb energy.}
\end{figure*}

The dielectric confinement exerted by the organic environment has been
known to influence the electronic properties of excitons in colloidal NPLs 
since early studies.\cite{achtstein2012electronic}
The low dielectric constant of the organic medium enables strong Coulomb
interactions, which give rise to large $E_b(X)$ values.
The influence on negative trions was recently calculated in Ref.~\citenum{ayari2020tuning}.
It was concluded that with increasing dielectric mismatch between the NPL and the surroundings,
$E_b(X^-)$ increases and its emission energy decreases,
all because of the weaker screening of electron-hole attraction.
The simulations however missed the self-energy correction $\Sigma$,
which accounts for the effect of the polarization induced by the carriers 
on themselves.

In Fig.~\ref{fig2} we extend the study of dielectric confinement by comparing 
neutral excitons with both negative and positive trions in CdSe and PbS NPLs,
as a function of the dielectric constant in the environment, $\varepsilon_{out}$. 
For all magnitudes under study, the reference energy is set to the case without
dielectric mismatch, $\varepsilon_{out}=\varepsilon_{NPL}$.
As can be seen, in spite of the Coulomb repulsions the behavior of X$^-$ (triangles)
and X$^+$ (squares) is similar to that of neutral excitons (dots). 
Namely, the net Coulomb interaction 
--i.e. the expectation value of attractions plus repulsions, not to be confused
with the binding energy -- (dashed line) 
becomes more and more attractive when $\varepsilon_{out}$ decreases. 
The enhancement of the attractions is however compensated by an increase of 
the self-energy correction, $\Sigma$ (dotted line). 
The latter term is greater than the former, which translates into an overall 
increase of the emission energy (solid line).
Similar findings have been observed for excitons in PbS\cite{yang2015electronic}
and CdSe\cite{rajadell2017excitons} NPLs.
The extension to trions implies that the emission wavelength in such species
can be also blueshifted by increasing the dielectric confinement.
To better see the magnitude of this effect, insets are included in Fig.~\ref{fig2},
which show that the blueshift can be few tens of meV large.
Energy shifts due to dielectric confinement have been proposed to be the driving 
mechanism for the formation of minibands in superlattices made of stacked NPLs.\cite{MovillaJPCL}
The results in Fig.~\ref{fig2} indicate that such bands could be built not only from
excitons but from trions as well.


\subsection{Balance between attractions and repulsions}

To visualize the relative strength of attractions and repulsions of trions in NPLs, 
we decompose the energy of excitons and trions as:
\begin{eqnarray}
E(X) &=& E_e + E_h + V_{eh}(X) + E_{gap}, \\
E(X^-) &=& 2E_e + E_h + 2V_{eh}(X^-) + V_{ee} + E_{gap}, \\
E(X^+) &=& E_e + 2E_h + 2V_{eh}(X^+) + V_{hh} + E_{gap}. 
\end{eqnarray}
\noindent where $E_e$ ($E_h$) is the expectation value of the electron (hole) 
single-particle terms in Hamiltonian ($\ref{eq:H}$), $V_{eh}$ that of the 
electron-hole Coulomb term and $V_{ee}$ ($V_{hh}$) that of electron-electron
(hole-hole) Coulomb term. 
\begin{figure*}[h!]
\includegraphics[scale=0.8]{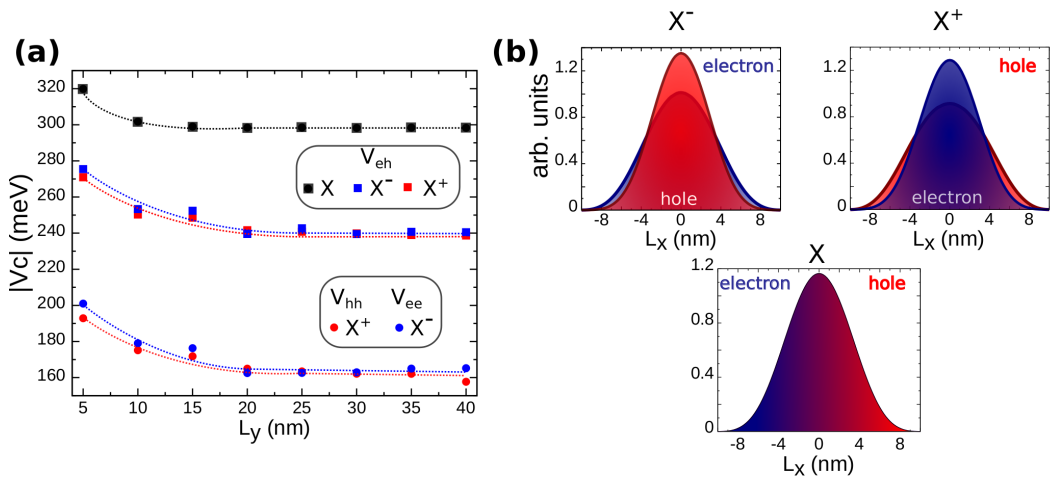}
\caption{\label{fig3} (a) Expectation value of the Coulomb terms of excitons
and trions, and (b) charge densities along the $x$ axis of the NPL
in a CdSe NPL with $L_x=20$ nm and $L_y=5$ nm.}
\end{figure*}
Figure \ref{fig3}(a) shows the calculated values of Coulomb energies
CdSe NPLs with varying lateral size $L_y$.
The electron-hole attraction of positive and negative trions is found to be similar, 
and clearly smaller than the attractions, $|V_{eh}(X^\pm)|=240-280$ meV,
while $|V_{eh}(X)|=300-320$ meV.
The repulsions are also similar for $X^-$ and $X^+$, but much smaller in energy
($160-200$ meV).
Altogether, these findings indicate that the strong electronic correlations
in colloidal NPLs make it possible for the trion to minimize repulsions and
maximize attractions, as attractive and repulsive terms would be identical 
in a strongly confined dot.\cite{lelong1996binding}
This explains the sizable binding energies of trions reported
in Fig.~\ref{fig1}.

The role of electronic correlations in deforming the wave functions
to stabilize excitonic states can be observed in Fig.~\ref{fig3}(b).
For a neutral exciton, electron and hole show identical charge densities
despite the different masses.
For trions, the repulsions between the carriers in excess 
(e.g.  electrons in X$^-$) lead to somewhat larger delocalization,
which reduces the electron-hole overlap substantially (note the
smaller height of the charge density in the central region).
These changes are reflected in the variational parameteres
related to effective Bohr radii: $r_B^X=2.23$ nm for the exciton,
$r_B^{X^-}=2.53$ nm and $r_B^{X^+}=2.98$ nm for the trions. 
The consequences on the optical properties properties are significant,
as we show in the next section.

\subsection{Emission spectra of CdSe NPLs}

\begin{figure*}[h!]
	\includegraphics[width=8cm]{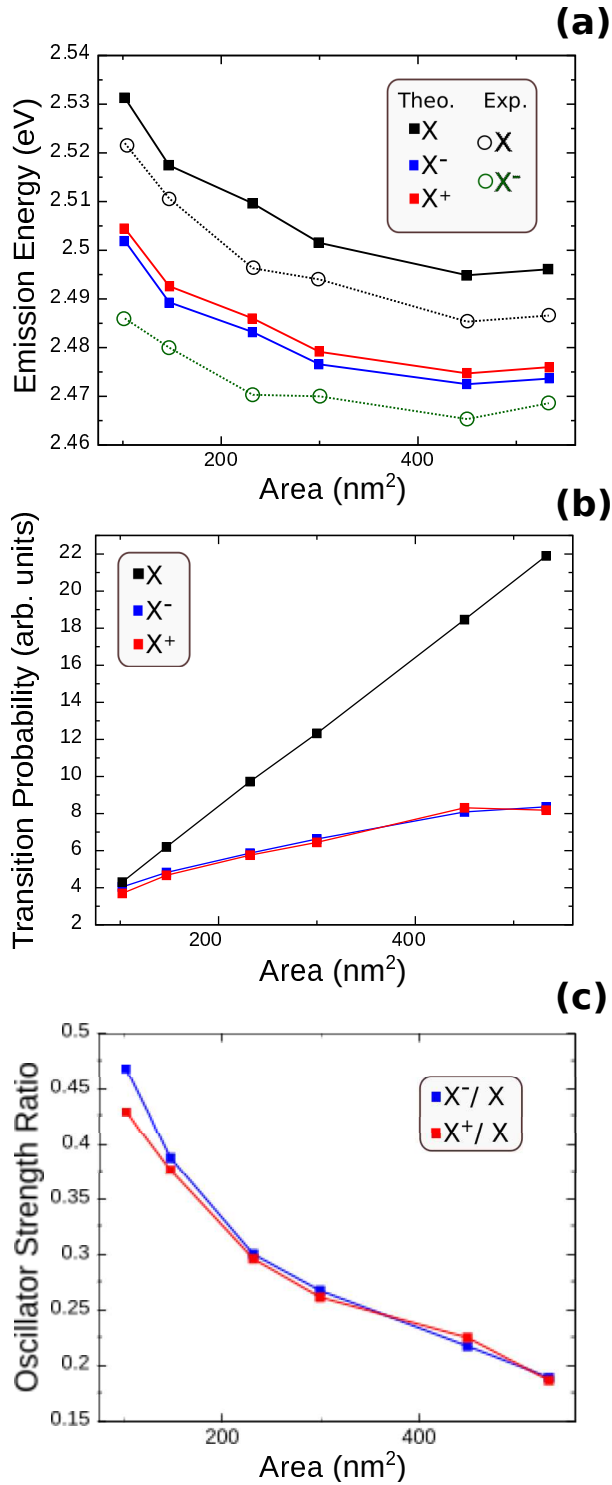}
\caption{\label{fig4} Analysis of emission properties for the CdSe NPLs of Ref.~\citenum{ayari2020tuning}.
(a) Comparison of theoretical and experimental emission energies. 
(b) Calculated recombination probability. (c) Calculated ratio between oscillator strengths of trions and excitons.
	We include positive trions for completeness.} 
\end{figure*}

Having studied the general behavior of trions in NPLs,
we next focus on the specific samples investigated in Ref.~\citenum{ayari2020tuning}
and analyze their emission spectra.
In Fig.~\ref{fig4}(a) we compare the calculated emission energies
of excitons and trions with those measured in the experiments.
 Our simulations overestimate the experimental values by $\sim 10$ meV,
but the overall agreement is good and, importantly, it is achieved with no fitting parameter.
The redshift of trions with respect to excitons ($20-30$ meV) is close to that of
the experiments ($20-35$ meV), which shows we obtain consistent binding energies.
The larger value as compared to CdSe/CdS/ZnS core/shell/shell NPLs ($E_b(X^-)=10.5$ meV)\cite{peng2020bright}
can be explained from: (i) the stronger dielectric confinement in core-only NPLs, 
which has significant influence on trion binding energies, 
as seen in the inset of Fig.~\ref{fig2}(a), and (ii) the quasi-type-II band 
alignment of CdSe/CdS, which promotes electron delocalization
and hence reduced Coulomb interaction with the hole.

Fig.~\ref{fig4}(b) compares the recombination probability of excitons and trions.
For the smallest NPLs (area $17\times 6$ nm$^2$), 
the probability is similar in all species. This is in spite of trions
having no spin-forbidden (dark) optical recombination channels. 
The reason is that the repulsions in trions reduce the electron-hole overlap,
as seen in the charge densities of Fig.~\ref{fig3}(b).
Consequently, the ratio of trion-to-exciton oscillator strength 
is about $0.45$ (see leftmost part of Fig.~\ref{fig4}(c)).
With increasing lateral dimensions of the NPL, the giant oscillator strength effect
boosts the recombination probability of X, which scales roughly linearly
with the area.\cite{feldmann1987linewidth,planelles2018tuning}
This is not observed in trions instead. 
For X$^\pm$, the increase is much weaker and saturates for large
areas ($> 400$ nm$^2$). 
 The resulting oscillator strength for large area NPLs is then 
smaller than that of excitons.
For the larger NPLs under consideration the ratio is $\sim 0.2$, 
(area $41\times13$ nm$^2$, see rightmost part of Fig.~\ref{fig4}(c)).
 
The oscillator strength ratios we obtain are in between those reported for CdSe/CdS/ZnS NPLs\cite{peng2020bright},
which are close to unity, and those calculated for CdSe NPLs in earlier studies,
 which range between $0.12$ and $0.04$.\cite{ayari2020tuning} 
 The smaller values as compared to CdSe/CdS/ZnS can be interpreted from the absence 
 of shells, which translates into larger dielectric mismatch.\cite{polovitsyn2017synthesis}
 This gives rise to  stronger electronic correlations, which in turn 
 permit shaping wave functions to minimize repulsions. 
 As for the larger values compared to Ref.~\citenum{ayari2020tuning} calculations,
 since both theoretical models rely on effective mass Hamiltonians, 
 describe dielectric confinement with methods of similar outcome (Fig.~\ref{fig1}),
 and provide good estimates of the experimentally observed emission and binding energies,
 the differences must be ascribed to the inclusion of many-body interactions in
 the wave function.
 Since VQMC calculations treat attractions and repulsions on equal footing, 
 the resulting charge distribution of trions is less likely to overestimate 
 the Bohr radius as compared to CI calculations.
 In this sense, it is not surprising that we obtain somewhat larger oscillator strengths.

 The saturation of the trion oscillator strength beyond 400 nm$^2$ in Fig.~\ref{fig4}(b)
 is also in contrast to the linear increase in the calculations of Ref.~\citenum{ayari2020tuning}.
 We note that our result is consistent with the expected behavior as one approaches the quantum well limit. 
 In the formation of a (say, negative) trion, the initial electron is originally delocalized 
 over the NPL volume, but it ends up being localized in the volume of the photocreated trion. 
 This localization leads to a reduction factor $1/A$ in the oscillator strength
 as compared to the exciton, with $A=L_x L_y$ the NPL area --
see for instance Eq. (13.51) in Ref.~\citenum{combescot2015excitons}--.
 Since for the exciton, $|\langle 0 | \delta_{\mathbf{r_e},\mathbf{r_h}} | \Psi_X \rangle|^2 \propto A$,\cite{feldmann1987linewidth,planelles2018tuning} 
 for the trion $\left|\langle \Psi_e | \delta_{\mathbf{r_{e_2}},\mathbf{r_{h_1}}} | \Psi_{X^-} \rangle \right|^2 \propto A^0$.
 Thus, for large NPLs we observe the 2D asymptotic behavior not only in the trion emission energies but 
 also in its oscillator strength.

%

\section{Conclusions}

We have compared exciton and trion optoelectronic properties in NPLs of CdSe and PbS.
The pronounced electronic correlations in colloidal NPLs allow trions to reduce 
Coulomb repulsions while keeping attractions strong. 
A few signatures of this are observed.
First, both negative and positive trions are redshifted with respect to the
exciton, with binding energies that exceed room temperature thermal energy in the 
case of CdSe with narrow lateral dimensions. 
 Second, contrary to excitons, the trion oscillator strength does not increase 
 linearly with the NPL area, as the giant oscillator strength effect is 
 significantly suppressed and vanishes for large areas.

Our calculations show that the description of dielectric confinement with
simple 2D Rytova-Keldysh potentials provide similar estimates of binding energies
to those of (fully 3D) image charge potentials in colloidal NPLs with typical 
(few-nm) thicknesses. However, they also evidence the need to add self-energy 
corrections so as to predict the right dependence of emission energy
on the dielectric mismatch.
These results are of interest for the design of optoelectronic devices
using homostructured NPLs populated with trions.

\begin{acknowledgement}
We are grateful to Sabrine Ayari and Sihem Jaziri for discussions.
Finacial support from MINECO project CTQ2017-83781-P (JP,JIC),
the European Research Council (ERC) via Consolidator Grant 724424-No-LIMIT
(D.M-P., I. M.-S.) and Generalitat Valenciana via Prometeo Grant Q-Devices
(Prometeo/2018/098) is gratefully acknowledged. 
\end{acknowledgement}





\bibliography{achemso}

\providecommand{\latin}[1]{#1}
\makeatletter
\providecommand{\doi}
  {\begingroup\let\do\@makeother\dospecials
  \catcode`\{=1 \catcode`\}=2 \doi@aux}
\providecommand{\doi@aux}[1]{\endgroup\texttt{#1}}
\makeatother
\providecommand*\mcitethebibliography{\thebibliography}
\csname @ifundefined\endcsname{endmcitethebibliography}
  {\let\endmcitethebibliography\endthebibliography}{}
\begin{mcitethebibliography}{67}
\providecommand*\natexlab[1]{#1}
\providecommand*\mciteSetBstSublistMode[1]{}
\providecommand*\mciteSetBstMaxWidthForm[2]{}
\providecommand*\mciteBstWouldAddEndPuncttrue
  {\def\EndOfBibitem{\unskip.}}
\providecommand*\mciteBstWouldAddEndPunctfalse
  {\let\EndOfBibitem\relax}
\providecommand*\mciteSetBstMidEndSepPunct[3]{}
\providecommand*\mciteSetBstSublistLabelBeginEnd[3]{}
\providecommand*\EndOfBibitem{}
\mciteSetBstSublistMode{f}
\mciteSetBstMaxWidthForm{subitem}{(\alph{mcitesubitemcount})}
\mciteSetBstSublistLabelBeginEnd
  {\mcitemaxwidthsubitemform\space}
  {\relax}
  {\relax}

\bibitem[Chen \latin{et~al.}(2014)Chen, Nadal, Mahler, Aubin, and
  Dubertret]{ZhoyingAFM}
Chen,~Z.; Nadal,~B.; Mahler,~B.; Aubin,~H.; Dubertret,~B. Quasi-2D Colloidal
  Semiconductor Nanoplatelets for Narrow Electroluminescence. \emph{Advanced
  Functional Materials} \textbf{2014}, \emph{24}, 295--302\relax
\mciteBstWouldAddEndPuncttrue
\mciteSetBstMidEndSepPunct{\mcitedefaultmidpunct}
{\mcitedefaultendpunct}{\mcitedefaultseppunct}\relax
\EndOfBibitem
\bibitem[Diroll(2020)]{diroll2020}
Diroll,~B.~T. Colloidal quantum wells for optoelectronic devices. \emph{Journal
  of Materials Chemistry C} \textbf{2020}, \emph{8}, 10628--10640\relax
\mciteBstWouldAddEndPuncttrue
\mciteSetBstMidEndSepPunct{\mcitedefaultmidpunct}
{\mcitedefaultendpunct}{\mcitedefaultseppunct}\relax
\EndOfBibitem
\bibitem[Dutta \latin{et~al.}(2021)Dutta, Medda, and Patra]{DuttaPCC}
Dutta,~A.; Medda,~A.; Patra,~A. Recent Advances and Perspectives on Colloidal
  Semiconductor Nanoplatelets for Optoelectronic Applications. \emph{The
  Journal of Physical Chemistry C} \textbf{2021}, \emph{125}, 20--30\relax
\mciteBstWouldAddEndPuncttrue
\mciteSetBstMidEndSepPunct{\mcitedefaultmidpunct}
{\mcitedefaultendpunct}{\mcitedefaultseppunct}\relax
\EndOfBibitem
\bibitem[Nasilowski \latin{et~al.}(2016)Nasilowski, Mahler, Lhuillier,
  Ithurria, and Dubertret]{nasilowski2016}
Nasilowski,~M.; Mahler,~B.; Lhuillier,~E.; Ithurria,~S.; Dubertret,~B.
  Two-dimensional colloidal nanocrystals. \emph{Chemical reviews}
  \textbf{2016}, \emph{116}, 10934--10982\relax
\mciteBstWouldAddEndPuncttrue
\mciteSetBstMidEndSepPunct{\mcitedefaultmidpunct}
{\mcitedefaultendpunct}{\mcitedefaultseppunct}\relax
\EndOfBibitem
\bibitem[Zhang \latin{et~al.}(2020)Zhang, Sun, Ye, Song, and
  Qu]{zhang2020heterostructures}
Zhang,~J.; Sun,~Y.; Ye,~S.; Song,~J.; Qu,~J. Heterostructures in
  Two-Dimensional CdSe Nanoplatelets: Synthesis, Optical Properties, and
  Applications. \emph{Chemistry of Materials} \textbf{2020}, \emph{32},
  9490--9507\relax
\mciteBstWouldAddEndPuncttrue
\mciteSetBstMidEndSepPunct{\mcitedefaultmidpunct}
{\mcitedefaultendpunct}{\mcitedefaultseppunct}\relax
\EndOfBibitem
\bibitem[Pun \latin{et~al.}(2021)Pun, Mazzotti, Mule, and
  Norris]{pun2021understanding}
Pun,~A.~B.; Mazzotti,~S.; Mule,~A.~S.; Norris,~D.~J. Understanding Discrete
  Growth in Semiconductor Nanocrystals: Nanoplatelets and Magic-Sized Clusters.
  \emph{Accounts of Chemical Research} \textbf{2021}, 743--748\relax
\mciteBstWouldAddEndPuncttrue
\mciteSetBstMidEndSepPunct{\mcitedefaultmidpunct}
{\mcitedefaultendpunct}{\mcitedefaultseppunct}\relax
\EndOfBibitem
\bibitem[Min \latin{et~al.}(2019)Min, Im, Hwang, Kim, Ahn, Choi, Hahn, Choi,
  Yoon, and Park]{min2019}
Min,~Y.; Im,~E.; Hwang,~G.-T.; Kim,~J.-W.; Ahn,~C.-W.; Choi,~J.-J.;
  Hahn,~B.-D.; Choi,~J.-H.; Yoon,~W.-H.; Park,~D.-S. Heterostructures in
  two-dimensional colloidal metal chalcogenides: Synthetic fundamentals and
  applications. \emph{Nano Research} \textbf{2019}, \emph{12}, 1750--1769\relax
\mciteBstWouldAddEndPuncttrue
\mciteSetBstMidEndSepPunct{\mcitedefaultmidpunct}
{\mcitedefaultendpunct}{\mcitedefaultseppunct}\relax
\EndOfBibitem
\bibitem[Ithurria \latin{et~al.}(2011)Ithurria, Tessier, Mahler, Lobo,
  Dubertret, and Efros]{ithurria2011colloidal}
Ithurria,~S.; Tessier,~M.; Mahler,~B.; Lobo,~R.; Dubertret,~B.; Efros,~A.~L.
  Colloidal nanoplatelets with two-dimensional electronic structure.
  \emph{Nature materials} \textbf{2011}, \emph{10}, 936--941\relax
\mciteBstWouldAddEndPuncttrue
\mciteSetBstMidEndSepPunct{\mcitedefaultmidpunct}
{\mcitedefaultendpunct}{\mcitedefaultseppunct}\relax
\EndOfBibitem
\bibitem[Achtstein \latin{et~al.}(2012)Achtstein, Schliwa, Prudnikau, Hardzei,
  Artemyev, Thomsen, and Woggon]{achtstein2012electronic}
Achtstein,~A.~W.; Schliwa,~A.; Prudnikau,~A.; Hardzei,~M.; Artemyev,~M.~V.;
  Thomsen,~C.; Woggon,~U. Electronic structure and exciton--phonon interaction
  in two-dimensional colloidal CdSe nanosheets. \emph{Nano letters}
  \textbf{2012}, \emph{12}, 3151--3157\relax
\mciteBstWouldAddEndPuncttrue
\mciteSetBstMidEndSepPunct{\mcitedefaultmidpunct}
{\mcitedefaultendpunct}{\mcitedefaultseppunct}\relax
\EndOfBibitem
\bibitem[Benchamekh \latin{et~al.}(2014)Benchamekh, Gippius, Even, Nestoklon,
  Jancu, Ithurria, Dubertret, Efros, and Voisin]{Benchamekh2014}
Benchamekh,~R.; Gippius,~N.~A.; Even,~J.; Nestoklon,~M.~O.; Jancu,~J.-M.;
  Ithurria,~S.; Dubertret,~B.; Efros,~A.~L.; Voisin,~P. Tight-binding
  calculations of image-charge effects in colloidal nanoscale platelets of
  CdSe. \emph{Phys. Rev. B} \textbf{2014}, \emph{89}, 035307\relax
\mciteBstWouldAddEndPuncttrue
\mciteSetBstMidEndSepPunct{\mcitedefaultmidpunct}
{\mcitedefaultendpunct}{\mcitedefaultseppunct}\relax
\EndOfBibitem
\bibitem[Yang and Wise(2015)Yang, and Wise]{yang2015electronic}
Yang,~J.; Wise,~F. Electronic states of lead-salt nanosheets. \emph{The Journal
  of Physical Chemistry C} \textbf{2015}, \emph{119}, 26809--26816\relax
\mciteBstWouldAddEndPuncttrue
\mciteSetBstMidEndSepPunct{\mcitedefaultmidpunct}
{\mcitedefaultendpunct}{\mcitedefaultseppunct}\relax
\EndOfBibitem
\bibitem[Bose \latin{et~al.}(2016)Bose, Song, Fan, and Zhang]{bose2016effect}
Bose,~S.; Song,~Z.; Fan,~W.~J.; Zhang,~D.~H. Effect of lateral size and
  thickness on the electronic structure and optical properties of quasi
  two-dimensional CdSe and CdS nanoplatelets. \emph{Journal of Applied Physics}
  \textbf{2016}, \emph{119}, 143107\relax
\mciteBstWouldAddEndPuncttrue
\mciteSetBstMidEndSepPunct{\mcitedefaultmidpunct}
{\mcitedefaultendpunct}{\mcitedefaultseppunct}\relax
\EndOfBibitem
\bibitem[Rajadell \latin{et~al.}(2017)Rajadell, Climente, and
  Planelles]{rajadell2017excitons}
Rajadell,~F.; Climente,~J.~I.; Planelles,~J. Excitons in core-only, core-shell
  and core-crown CdSe nanoplatelets: Interplay between in-plane electron-hole
  correlation, spatial confinement, and dielectric confinement. \emph{Physical
  Review B} \textbf{2017}, \emph{96}, 035307\relax
\mciteBstWouldAddEndPuncttrue
\mciteSetBstMidEndSepPunct{\mcitedefaultmidpunct}
{\mcitedefaultendpunct}{\mcitedefaultseppunct}\relax
\EndOfBibitem
\bibitem[Richter(2017)]{richter2017nanoplatelets}
Richter,~M. Nanoplatelets as material system between strong confinement and
  weak confinement. \emph{Physical Review Materials} \textbf{2017}, \emph{1},
  016001\relax
\mciteBstWouldAddEndPuncttrue
\mciteSetBstMidEndSepPunct{\mcitedefaultmidpunct}
{\mcitedefaultendpunct}{\mcitedefaultseppunct}\relax
\EndOfBibitem
\bibitem[Polovitsyn \latin{et~al.}(2017)Polovitsyn, Dang, Movilla,
  Mart{\'\i}n-Garc{\'\i}a, Khan, Bertrand, Brescia, and
  Moreels]{polovitsyn2017synthesis}
Polovitsyn,~A.; Dang,~Z.; Movilla,~J.~L.; Mart{\'\i}n-Garc{\'\i}a,~B.;
  Khan,~A.~H.; Bertrand,~G.~H.; Brescia,~R.; Moreels,~I. Synthesis of
  air-stable CdSe/ZnS core--shell nanoplatelets with tunable emission
  wavelength. \emph{Chemistry of Materials} \textbf{2017}, \emph{29},
  5671--5680\relax
\mciteBstWouldAddEndPuncttrue
\mciteSetBstMidEndSepPunct{\mcitedefaultmidpunct}
{\mcitedefaultendpunct}{\mcitedefaultseppunct}\relax
\EndOfBibitem
\bibitem[Chu \latin{et~al.}(2018)Chu, Livache, Ithurria, and
  Lhuillier]{chu2018electronic}
Chu,~A.; Livache,~C.; Ithurria,~S.; Lhuillier,~E. Electronic structure
  robustness and design rules for 2D colloidal heterostructures. \emph{Journal
  of Applied Physics} \textbf{2018}, \emph{123}, 035701\relax
\mciteBstWouldAddEndPuncttrue
\mciteSetBstMidEndSepPunct{\mcitedefaultmidpunct}
{\mcitedefaultendpunct}{\mcitedefaultseppunct}\relax
\EndOfBibitem
\bibitem[Achtstein \latin{et~al.}(2018)Achtstein, Marquardt, Scott, Ibrahim,
  Riedl, Prudnikau, Antanovich, Owschimikow, Lindner, and
  Artemyev]{achtstein2018impact}
Achtstein,~A.~W.; Marquardt,~O.; Scott,~R.; Ibrahim,~M.; Riedl,~T.;
  Prudnikau,~A.~V.; Antanovich,~A.; Owschimikow,~N.; Lindner,~J.~K.;
  Artemyev,~M. Impact of shell growth on recombination dynamics and
  exciton--phonon interaction in CdSe--CdS core--shell nanoplatelets. \emph{ACS
  nano} \textbf{2018}, \emph{12}, 9476--9483\relax
\mciteBstWouldAddEndPuncttrue
\mciteSetBstMidEndSepPunct{\mcitedefaultmidpunct}
{\mcitedefaultendpunct}{\mcitedefaultseppunct}\relax
\EndOfBibitem
\bibitem[Specht \latin{et~al.}(2019)Specht, Scott, Castro, Christodoulou,
  Bertrand, Prudnikau, Antanovich, Siebbeles, Owschimikow, and
  Moreels]{specht2019size}
Specht,~J.~F.; Scott,~R.; Castro,~M.~C.; Christodoulou,~S.; Bertrand,~G.~H.;
  Prudnikau,~A.~V.; Antanovich,~A.; Siebbeles,~L.~D.; Owschimikow,~N.;
  Moreels,~I. Size-dependent exciton substructure in CdSe nanoplatelets and its
  relation to photoluminescence dynamics. \emph{Nanoscale} \textbf{2019},
  \emph{11}, 12230--12241\relax
\mciteBstWouldAddEndPuncttrue
\mciteSetBstMidEndSepPunct{\mcitedefaultmidpunct}
{\mcitedefaultendpunct}{\mcitedefaultseppunct}\relax
\EndOfBibitem
\bibitem[Llusar \latin{et~al.}(2019)Llusar, Planelles, and
  Climente]{llusar2019strain}
Llusar,~J.; Planelles,~J.; Climente,~J.~I. Strain in lattice-mismatched
  CdSe-based core/shell nanoplatelets. \emph{The Journal of Physical Chemistry
  C} \textbf{2019}, \emph{123}, 21299--21306\relax
\mciteBstWouldAddEndPuncttrue
\mciteSetBstMidEndSepPunct{\mcitedefaultmidpunct}
{\mcitedefaultendpunct}{\mcitedefaultseppunct}\relax
\EndOfBibitem
\bibitem[Steinmetz \latin{et~al.}(2020)Steinmetz, Climente, Pandya, Planelles,
  Margaillan, Puttisong, Dufour, Ithurria, Sharma, and
  Lakhwani]{steinmetz2020emission}
Steinmetz,~V.; Climente,~J.~I.; Pandya,~R.; Planelles,~J.; Margaillan,~F.;
  Puttisong,~Y.; Dufour,~M.; Ithurria,~S.; Sharma,~A.; Lakhwani,~G. Emission
  State Structure and Linewidth Broadening Mechanisms in Type-II CdSe/CdTe
  Core--Crown Nanoplatelets: A Combined Theoretical--Single Nanocrystal Optical
  Study. \emph{The Journal of Physical Chemistry C} \textbf{2020}, \emph{124},
  17352--17363\relax
\mciteBstWouldAddEndPuncttrue
\mciteSetBstMidEndSepPunct{\mcitedefaultmidpunct}
{\mcitedefaultendpunct}{\mcitedefaultseppunct}\relax
\EndOfBibitem
\bibitem[Macias-Pinilla \latin{et~al.}(2021)Macias-Pinilla,
  Echeverr{\'i}a-Arrondo, Gualdr{\'o}n~Reyes, Agouram, Mu{\~n}oz-Sanjos{\'e},
  Planelles, Mora-Ser{\'o}, and Climente]{macias2021morphology}
Macias-Pinilla,~D.~F.; Echeverr{\'i}a-Arrondo,~C.; Gualdr{\'o}n~Reyes,~A.~F.;
  Agouram,~S.; Mu{\~n}oz-Sanjos{\'e},~V.; Planelles,~J.; Mora-Ser{\'o},~I.;
  Climente,~J.~I. Morphology and Band Structure of Orthorhombic PbS
  Nanoplatelets: An Indirect Band Gap Material. \emph{Chemistry of Materials}
  \textbf{2021}, \emph{33}, 420--429\relax
\mciteBstWouldAddEndPuncttrue
\mciteSetBstMidEndSepPunct{\mcitedefaultmidpunct}
{\mcitedefaultendpunct}{\mcitedefaultseppunct}\relax
\EndOfBibitem
\bibitem[Greenwood \latin{et~al.}(2021)Greenwood, Mazzotti, Norris, and
  Galli]{Greenwood2021}
Greenwood,~A.~R.; Mazzotti,~S.; Norris,~D.~J.; Galli,~G. Determining the
  Structure–Property Relationships of Quasi-Two-Dimensional Semiconductor
  Nanoplatelets. \emph{The Journal of Physical Chemistry C} \textbf{2021},
  \emph{125}, 4820--4827\relax
\mciteBstWouldAddEndPuncttrue
\mciteSetBstMidEndSepPunct{\mcitedefaultmidpunct}
{\mcitedefaultendpunct}{\mcitedefaultseppunct}\relax
\EndOfBibitem
\bibitem[Lorenzon \latin{et~al.}(2015)Lorenzon, Christodoulou, Vaccaro,
  Pedrini, Meinardi, Moreels, and Brovelli]{Lorenzon2015}
Lorenzon,~M.; Christodoulou,~S.; Vaccaro,~G.; Pedrini,~J.; Meinardi,~F.;
  Moreels,~I.; Brovelli,~S. Reversed oxygen sensing using colloidal quantum
  wells towards highly emissive photoresponsive varnishes. \emph{Nature
  Communications} \textbf{2015}, \emph{6}, 6434\relax
\mciteBstWouldAddEndPuncttrue
\mciteSetBstMidEndSepPunct{\mcitedefaultmidpunct}
{\mcitedefaultendpunct}{\mcitedefaultseppunct}\relax
\EndOfBibitem
\bibitem[Yu \latin{et~al.}(2019)Yu, Zhang, Pang, Sun, and Chen]{yu2019effect}
Yu,~J.; Zhang,~C.; Pang,~G.; Sun,~X.~W.; Chen,~R. Effect of lateral size and
  surface passivation on the near-band-edge excitonic emission from
  quasi-two-dimensional CdSe nanoplatelets. \emph{ACS applied materials \&
  interfaces} \textbf{2019}, \emph{11}, 41821--41827\relax
\mciteBstWouldAddEndPuncttrue
\mciteSetBstMidEndSepPunct{\mcitedefaultmidpunct}
{\mcitedefaultendpunct}{\mcitedefaultseppunct}\relax
\EndOfBibitem
\bibitem[Shornikova \latin{et~al.}(2020)Shornikova, Yakovlev, Biadala, Crooker,
  Belykh, Kochiev, Kuntzmann, Nasilowski, Dubertret, and Bayer]{Shornikova2020}
Shornikova,~E.~V.; Yakovlev,~D.~R.; Biadala,~L.; Crooker,~S.~A.; Belykh,~V.~V.;
  Kochiev,~M.~V.; Kuntzmann,~A.; Nasilowski,~M.; Dubertret,~B.; Bayer,~M.
  Negatively Charged Excitons in CdSe Nanoplatelets. \emph{Nano Letters}
  \textbf{2020}, \emph{20}, 1370--1377, PMID: 31960677\relax
\mciteBstWouldAddEndPuncttrue
\mciteSetBstMidEndSepPunct{\mcitedefaultmidpunct}
{\mcitedefaultendpunct}{\mcitedefaultseppunct}\relax
\EndOfBibitem
\bibitem[Antolinez \latin{et~al.}(2019)Antolinez, Rabouw, Rossinelli, Cui, and
  Norris]{antolinez2019observation}
Antolinez,~F.~V.; Rabouw,~F.~T.; Rossinelli,~A.~A.; Cui,~J.; Norris,~D.~J.
  Observation of electron shakeup in CdSe/CdS core/shell nanoplatelets.
  \emph{Nano letters} \textbf{2019}, \emph{19}, 8495--8502\relax
\mciteBstWouldAddEndPuncttrue
\mciteSetBstMidEndSepPunct{\mcitedefaultmidpunct}
{\mcitedefaultendpunct}{\mcitedefaultseppunct}\relax
\EndOfBibitem
\bibitem[Antolinez \latin{et~al.}(2020)Antolinez, Rabouw, Rossinelli, Keitel,
  Cocina, Becker, and Norris]{antolinez2020trion}
Antolinez,~F.~V.; Rabouw,~F.~T.; Rossinelli,~A.~A.; Keitel,~R.~C.; Cocina,~A.;
  Becker,~M.~A.; Norris,~D.~J. Trion emission dominates the low-temperature
  photoluminescence of CdSe nanoplatelets. \emph{Nano Letters} \textbf{2020},
  \emph{20}, 5814--5820\relax
\mciteBstWouldAddEndPuncttrue
\mciteSetBstMidEndSepPunct{\mcitedefaultmidpunct}
{\mcitedefaultendpunct}{\mcitedefaultseppunct}\relax
\EndOfBibitem
\bibitem[Peng \latin{et~al.}(2020)Peng, Otten, Hazarika, Coropceanu, Cygorek,
  Wiederrecht, Hawrylak, Talapin, and Ma]{peng2020bright}
Peng,~L.; Otten,~M.; Hazarika,~A.; Coropceanu,~I.; Cygorek,~M.;
  Wiederrecht,~G.~P.; Hawrylak,~P.; Talapin,~D.~V.; Ma,~X. Bright trion
  emission from semiconductor nanoplatelets. \emph{Physical Review Materials}
  \textbf{2020}, \emph{4}, 056006\relax
\mciteBstWouldAddEndPuncttrue
\mciteSetBstMidEndSepPunct{\mcitedefaultmidpunct}
{\mcitedefaultendpunct}{\mcitedefaultseppunct}\relax
\EndOfBibitem
\bibitem[Ayari \latin{et~al.}(2020)Ayari, Quick, Owschimikow, Christodoulou,
  Bertrand, Artemyev, Moreels, Woggon, Jaziri, and Achtstein]{ayari2020tuning}
Ayari,~S.; Quick,~M.~T.; Owschimikow,~N.; Christodoulou,~S.; Bertrand,~G.~H.;
  Artemyev,~M.; Moreels,~I.; Woggon,~U.; Jaziri,~S.; Achtstein,~A.~W. Tuning
  trion binding energy and oscillator strength in a laterally finite 2D system:
  CdSe nanoplatelets as a model system for trion properties. \emph{Nanoscale}
  \textbf{2020}, \emph{12}, 14448--14458\relax
\mciteBstWouldAddEndPuncttrue
\mciteSetBstMidEndSepPunct{\mcitedefaultmidpunct}
{\mcitedefaultendpunct}{\mcitedefaultseppunct}\relax
\EndOfBibitem
\bibitem[Kunneman \latin{et~al.}(2013)Kunneman, Tessier, Heuclin, Dubertret,
  Aulin, Grozema, Schins, and Siebbeles]{kunneman2013}
Kunneman,~L.~T.; Tessier,~M.~D.; Heuclin,~H.; Dubertret,~B.; Aulin,~Y.~V.;
  Grozema,~F.~C.; Schins,~J.~M.; Siebbeles,~L.~D. Bimolecular Auger
  recombination of electron--hole pairs in two-dimensional CdSe and CdSe/CdZnS
  core/shell nanoplatelets. \emph{The Journal of Physical Chemistry Letters}
  \textbf{2013}, \emph{4}, 3574--3578\relax
\mciteBstWouldAddEndPuncttrue
\mciteSetBstMidEndSepPunct{\mcitedefaultmidpunct}
{\mcitedefaultendpunct}{\mcitedefaultseppunct}\relax
\EndOfBibitem
\bibitem[Baghani \latin{et~al.}(2015)Baghani, O’Leary, Fedin, Talapin, and
  Pelton]{baghani2015}
Baghani,~E.; O’Leary,~S.~K.; Fedin,~I.; Talapin,~D.~V.; Pelton,~M.
  Auger-limited carrier recombination and relaxation in CdSe colloidal quantum
  wells. \emph{The journal of physical chemistry letters} \textbf{2015},
  \emph{6}, 1032--1036\relax
\mciteBstWouldAddEndPuncttrue
\mciteSetBstMidEndSepPunct{\mcitedefaultmidpunct}
{\mcitedefaultendpunct}{\mcitedefaultseppunct}\relax
\EndOfBibitem
\bibitem[Li and Lian(2017)Li, and Lian]{li2017}
Li,~Q.; Lian,~T. Area-and thickness-dependent biexciton Auger recombination in
  colloidal CdSe nanoplatelets: breaking the “Universal Volume Scaling
  Law”. \emph{Nano letters} \textbf{2017}, \emph{17}, 3152--3158\relax
\mciteBstWouldAddEndPuncttrue
\mciteSetBstMidEndSepPunct{\mcitedefaultmidpunct}
{\mcitedefaultendpunct}{\mcitedefaultseppunct}\relax
\EndOfBibitem
\bibitem[Philbin \latin{et~al.}(2020)Philbin, Brumberg, Diroll, Cho, Talapin,
  Schaller, and Rabani]{philbin2020}
Philbin,~J.~P.; Brumberg,~A.; Diroll,~B.~T.; Cho,~W.; Talapin,~D.~V.;
  Schaller,~R.~D.; Rabani,~E. Area and thickness dependence of Auger
  recombination in nanoplatelets. \emph{The Journal of Chemical Physics}
  \textbf{2020}, \emph{153}, 054104\relax
\mciteBstWouldAddEndPuncttrue
\mciteSetBstMidEndSepPunct{\mcitedefaultmidpunct}
{\mcitedefaultendpunct}{\mcitedefaultseppunct}\relax
\EndOfBibitem
\bibitem[L{\"o}bl \latin{et~al.}(2020)L{\"o}bl, Spinnler, Javadi, Zhai, Nguyen,
  Ritzmann, Midolo, Lodahl, Wieck, and Ludwig]{lobl2020radiative}
L{\"o}bl,~M.~C.; Spinnler,~C.; Javadi,~A.; Zhai,~L.; Nguyen,~G.~N.;
  Ritzmann,~J.; Midolo,~L.; Lodahl,~P.; Wieck,~A.~D.; Ludwig,~A. Radiative
  Auger process in the single-photon limit. \emph{Nature Nanotechnology}
  \textbf{2020}, \emph{15}, 558--562\relax
\mciteBstWouldAddEndPuncttrue
\mciteSetBstMidEndSepPunct{\mcitedefaultmidpunct}
{\mcitedefaultendpunct}{\mcitedefaultseppunct}\relax
\EndOfBibitem
\bibitem[Llusar and Climente(2020)Llusar, and Climente]{llusar2020nature}
Llusar,~J.; Climente,~J.~I. Nature and Control of Shakeup Processes in
  Colloidal Nanoplatelets. \emph{ACS Photonics} \textbf{2020}, \emph{7},
  3086--3095\relax
\mciteBstWouldAddEndPuncttrue
\mciteSetBstMidEndSepPunct{\mcitedefaultmidpunct}
{\mcitedefaultendpunct}{\mcitedefaultseppunct}\relax
\EndOfBibitem
\bibitem[Brumberg \latin{et~al.}(2019)Brumberg, Harvey, Philbin, Diroll, Lee,
  Crooker, Wasielewski, Rabani, and Schaller]{brumberg2019determination}
Brumberg,~A.; Harvey,~S.~M.; Philbin,~J.~P.; Diroll,~B.~T.; Lee,~B.;
  Crooker,~S.~A.; Wasielewski,~M.~R.; Rabani,~E.; Schaller,~R.~D. Determination
  of the In-Plane Exciton Radius in 2D CdSe Nanoplatelets via Magneto-optical
  Spectroscopy. \emph{ACS nano} \textbf{2019}, \emph{13}, 8589--8596\relax
\mciteBstWouldAddEndPuncttrue
\mciteSetBstMidEndSepPunct{\mcitedefaultmidpunct}
{\mcitedefaultendpunct}{\mcitedefaultseppunct}\relax
\EndOfBibitem
\bibitem[Planelles(2017)]{planelles2017simple}
Planelles,~J. Simple correlated wave-function for excitons in 0D, quasi-1D and
  quasi-2D quantum dots. \emph{Theoretical Chemistry Accounts} \textbf{2017},
  \emph{136}, 1--16\relax
\mciteBstWouldAddEndPuncttrue
\mciteSetBstMidEndSepPunct{\mcitedefaultmidpunct}
{\mcitedefaultendpunct}{\mcitedefaultseppunct}\relax
\EndOfBibitem
\bibitem[Rontani \latin{et~al.}(2017)Rontani, Eriksson, {\AA}berg, and
  Reimann]{Rontani_2017}
Rontani,~M.; Eriksson,~G.; {\AA}berg,~S.; Reimann,~S.~M. On the renormalization
  of contact interactions for the configuration-interaction method in
  two-dimensions. \emph{Journal of Physics B: Atomic, Molecular and Optical
  Physics} \textbf{2017}, \emph{50}, 065301\relax
\mciteBstWouldAddEndPuncttrue
\mciteSetBstMidEndSepPunct{\mcitedefaultmidpunct}
{\mcitedefaultendpunct}{\mcitedefaultseppunct}\relax
\EndOfBibitem
\bibitem[Planelles and Climente(2021)Planelles, and
  Climente]{planelles2021simple}
Planelles,~J.; Climente,~J.~I. A simple variational quantum Monte
  Carlo-effective mass approach for excitons and trions in quantum dots.
  \emph{Computer Physics Communications} \textbf{2021}, \emph{261},
  107782\relax
\mciteBstWouldAddEndPuncttrue
\mciteSetBstMidEndSepPunct{\mcitedefaultmidpunct}
{\mcitedefaultendpunct}{\mcitedefaultseppunct}\relax
\EndOfBibitem
\bibitem[Carey \latin{et~al.}(2015)Carey, Abdelhady, Ning, Thon, Bakr, and
  Sargent]{carey2015colloidal}
Carey,~G.~H.; Abdelhady,~A.~L.; Ning,~Z.; Thon,~S.~M.; Bakr,~O.~M.;
  Sargent,~E.~H. Colloidal quantum dot solar cells. \emph{Chemical reviews}
  \textbf{2015}, \emph{115}, 12732--12763\relax
\mciteBstWouldAddEndPuncttrue
\mciteSetBstMidEndSepPunct{\mcitedefaultmidpunct}
{\mcitedefaultendpunct}{\mcitedefaultseppunct}\relax
\EndOfBibitem
\bibitem[S{\'a}nchez-Godoy \latin{et~al.}(2020)S{\'a}nchez-Godoy, Erazo,
  Gualdr{\'o}n-Reyes, Khan, Agouram, Barea, Rodriguez, Zaraz{\'u}a, Ortiz,
  Cort{\'e}s, \latin{et~al.} others]{sanchez2020preferred}
S{\'a}nchez-Godoy,~H.~E.; Erazo,~E.~A.; Gualdr{\'o}n-Reyes,~A.~F.; Khan,~A.~H.;
  Agouram,~S.; Barea,~E.~M.; Rodriguez,~R.~A.; Zaraz{\'u}a,~I.; Ortiz,~P.;
  Cort{\'e}s,~M.~T. \latin{et~al.}  Preferred Growth Direction by PbS
  Nanoplatelets Preserves Perovskite Infrared Light Harvesting for Stable,
  Reproducible, and Efficient Solar Cells. \emph{Advanced Energy Materials}
  \textbf{2020}, \emph{10}, 2002422\relax
\mciteBstWouldAddEndPuncttrue
\mciteSetBstMidEndSepPunct{\mcitedefaultmidpunct}
{\mcitedefaultendpunct}{\mcitedefaultseppunct}\relax
\EndOfBibitem
\bibitem[De~Iacovo \latin{et~al.}(2016)De~Iacovo, Venettacci, Colace, Scopa,
  and Foglia]{DeIacovo2016}
De~Iacovo,~A.; Venettacci,~C.; Colace,~L.; Scopa,~L.; Foglia,~S. PbS Colloidal
  Quantum Dot Photodetectors operating in the near infrared. \emph{Scientific
  Reports} \textbf{2016}, \emph{6}, 37913\relax
\mciteBstWouldAddEndPuncttrue
\mciteSetBstMidEndSepPunct{\mcitedefaultmidpunct}
{\mcitedefaultendpunct}{\mcitedefaultseppunct}\relax
\EndOfBibitem
\bibitem[Sanchez \latin{et~al.}(2014)Sanchez, Binetti, Torre, Garcia-Belmonte,
  Striccoli, and Mora-Sero]{sanchez2014all}
Sanchez,~R.~S.; Binetti,~E.; Torre,~J.~A.; Garcia-Belmonte,~G.; Striccoli,~M.;
  Mora-Sero,~I. All solution processed low turn-on voltage near infrared LEDs
  based on core--shell PbS--CdS quantum dots with inverted device structure.
  \emph{Nanoscale} \textbf{2014}, \emph{6}, 8551--8555\relax
\mciteBstWouldAddEndPuncttrue
\mciteSetBstMidEndSepPunct{\mcitedefaultmidpunct}
{\mcitedefaultendpunct}{\mcitedefaultseppunct}\relax
\EndOfBibitem
\bibitem[Fan \latin{et~al.}(2015)Fan, Kanjanaboos, Saravanapavanantham,
  Beauregard, Ingram, Yassitepe, Adachi, Voznyy, Johnston, Walters,
  \latin{et~al.} others]{fan2015colloidal}
Fan,~F.; Kanjanaboos,~P.; Saravanapavanantham,~M.; Beauregard,~E.; Ingram,~G.;
  Yassitepe,~E.; Adachi,~M.~M.; Voznyy,~O.; Johnston,~A.~K.; Walters,~G.
  \latin{et~al.}  Colloidal CdSe1--x S x Nanoplatelets with Narrow and
  Continuously-Tunable Electroluminescence. \emph{Nano letters} \textbf{2015},
  \emph{15}, 4611--4615\relax
\mciteBstWouldAddEndPuncttrue
\mciteSetBstMidEndSepPunct{\mcitedefaultmidpunct}
{\mcitedefaultendpunct}{\mcitedefaultseppunct}\relax
\EndOfBibitem
\bibitem[van Der~Bok \latin{et~al.}(2020)van Der~Bok, Dekker, Peerlings,
  Salzmann, and Meijerink]{van2020luminescence}
van Der~Bok,~J.~C.; Dekker,~D.~M.; Peerlings,~M.~L.; Salzmann,~B.~B.;
  Meijerink,~A. Luminescence Line Broadening of CdSe Nanoplatelets and Quantum
  Dots for Application in w-LEDs. \emph{The Journal of Physical Chemistry C}
  \textbf{2020}, \emph{124}, 12153--12160\relax
\mciteBstWouldAddEndPuncttrue
\mciteSetBstMidEndSepPunct{\mcitedefaultmidpunct}
{\mcitedefaultendpunct}{\mcitedefaultseppunct}\relax
\EndOfBibitem
\bibitem[Guzelturk \latin{et~al.}(2014)Guzelturk, Kelestemur, Olutas,
  Delikanli, and Demir]{guzelturk2014amplified}
Guzelturk,~B.; Kelestemur,~Y.; Olutas,~M.; Delikanli,~S.; Demir,~H.~V.
  Amplified spontaneous emission and lasing in colloidal nanoplatelets.
  \emph{ACS nano} \textbf{2014}, \emph{8}, 6599--6605\relax
\mciteBstWouldAddEndPuncttrue
\mciteSetBstMidEndSepPunct{\mcitedefaultmidpunct}
{\mcitedefaultendpunct}{\mcitedefaultseppunct}\relax
\EndOfBibitem
\bibitem[Kumagai and Takagahara(1989)Kumagai, and
  Takagahara]{kumagai1989excitonic}
Kumagai,~M.; Takagahara,~T. Excitonic and nonlinear-optical properties of
  dielectric quantum-well structures. \emph{Physical Review B} \textbf{1989},
  \emph{40}, 12359\relax
\mciteBstWouldAddEndPuncttrue
\mciteSetBstMidEndSepPunct{\mcitedefaultmidpunct}
{\mcitedefaultendpunct}{\mcitedefaultseppunct}\relax
\EndOfBibitem
\bibitem[Cudazzo \latin{et~al.}(2011)Cudazzo, Tokatly, and
  Rubio]{cudazzo2011dielectric}
Cudazzo,~P.; Tokatly,~I.~V.; Rubio,~A. Dielectric screening in two-dimensional
  insulators: Implications for excitonic and impurity states in graphane.
  \emph{Physical Review B} \textbf{2011}, \emph{84}, 085406\relax
\mciteBstWouldAddEndPuncttrue
\mciteSetBstMidEndSepPunct{\mcitedefaultmidpunct}
{\mcitedefaultendpunct}{\mcitedefaultseppunct}\relax
\EndOfBibitem
\bibitem[Jacak \latin{et~al.}(1998)Jacak, Hawrylak, and Wojs]{Pawel_book}
Jacak,~L.; Hawrylak,~P.; Wojs,~A. \emph{Quantum Dots}; Springer, 1998\relax
\mciteBstWouldAddEndPuncttrue
\mciteSetBstMidEndSepPunct{\mcitedefaultmidpunct}
{\mcitedefaultendpunct}{\mcitedefaultseppunct}\relax
\EndOfBibitem
\bibitem[Antu \latin{et~al.}(2018)Antu, Jiang, Premathilka, Tang, Hu, Roy, and
  Sun]{antu2018bright}
Antu,~A.~D.; Jiang,~Z.; Premathilka,~S.~M.; Tang,~Y.; Hu,~J.; Roy,~A.; Sun,~L.
  Bright colloidal PbS nanoribbons. \emph{Chemistry of Materials}
  \textbf{2018}, \emph{30}, 3697--3703\relax
\mciteBstWouldAddEndPuncttrue
\mciteSetBstMidEndSepPunct{\mcitedefaultmidpunct}
{\mcitedefaultendpunct}{\mcitedefaultseppunct}\relax
\EndOfBibitem
\bibitem[L{\"u}dde(1959)]{ludde1959dielektrische}
L{\"u}dde,~K. Das dielektrische und refraktodensimetrische Verhalten
  fl{\"u}ssiger Fette beim Altern. \emph{Fette, Seifen, Anstrichmittel}
  \textbf{1959}, \emph{61}, 1156--1163\relax
\mciteBstWouldAddEndPuncttrue
\mciteSetBstMidEndSepPunct{\mcitedefaultmidpunct}
{\mcitedefaultendpunct}{\mcitedefaultseppunct}\relax
\EndOfBibitem
\bibitem[Adachi(2004)]{adachi2004handbook}
Adachi,~S. \emph{Handbook on physical properties of semiconductors}; Springer
  Science \& Business Media, 2004; Vol.~2\relax
\mciteBstWouldAddEndPuncttrue
\mciteSetBstMidEndSepPunct{\mcitedefaultmidpunct}
{\mcitedefaultendpunct}{\mcitedefaultseppunct}\relax
\EndOfBibitem
\bibitem[Gupta and Kumar(1983)Gupta, and Kumar]{gupta1983analysis}
Gupta,~B.; Kumar,~V. Analysis of effective compressibilities in PbS, PbSe, PbTe
  and SnTe. \emph{Solid State Communications} \textbf{1983}, \emph{45},
  745--747\relax
\mciteBstWouldAddEndPuncttrue
\mciteSetBstMidEndSepPunct{\mcitedefaultmidpunct}
{\mcitedefaultendpunct}{\mcitedefaultseppunct}\relax
\EndOfBibitem
\bibitem[Kang and Wise(1997)Kang, and Wise]{kang1997electronic}
Kang,~I.; Wise,~F.~W. Electronic structure and optical properties of PbS and
  PbSe quantum dots. \emph{JOSA B} \textbf{1997}, \emph{14}, 1632--1646\relax
\mciteBstWouldAddEndPuncttrue
\mciteSetBstMidEndSepPunct{\mcitedefaultmidpunct}
{\mcitedefaultendpunct}{\mcitedefaultseppunct}\relax
\EndOfBibitem
\bibitem[Hoda and Chang(1975)Hoda, and Chang]{hoda1975phase}
Hoda,~S.~N.; Chang,~L.~L. Phase relations in the systems PbS-Ag2S-Sb2S3 and
  PbS-Ag2S-Bi2S3. \emph{American Mineralogist: Journal of Earth and Planetary
  Materials} \textbf{1975}, \emph{60}, 621--633\relax
\mciteBstWouldAddEndPuncttrue
\mciteSetBstMidEndSepPunct{\mcitedefaultmidpunct}
{\mcitedefaultendpunct}{\mcitedefaultseppunct}\relax
\EndOfBibitem
\bibitem[Zelewski \latin{et~al.}(2019)Zelewski, Nawrot, Zak, Gladysiewicz, Nyk,
  and Kudrawiec]{zelewski2019exciton}
Zelewski,~S.~J.; Nawrot,~K.~C.; Zak,~A.; Gladysiewicz,~M.; Nyk,~M.;
  Kudrawiec,~R. Exciton binding energy of two-dimensional highly luminescent
  colloidal nanostructures determined from combined optical and photoacoustic
  spectroscopies. \emph{The journal of physical chemistry letters}
  \textbf{2019}, \emph{10}, 3459--3464\relax
\mciteBstWouldAddEndPuncttrue
\mciteSetBstMidEndSepPunct{\mcitedefaultmidpunct}
{\mcitedefaultendpunct}{\mcitedefaultseppunct}\relax
\EndOfBibitem
\bibitem[Khan \latin{et~al.}(2017)Khan, Brescia, Polovitsyn, Angeloni,
  Mart{\'\i}n-Garc{\'\i}a, and Moreels]{khan2017near}
Khan,~A.~H.; Brescia,~R.; Polovitsyn,~A.; Angeloni,~I.;
  Mart{\'\i}n-Garc{\'\i}a,~B.; Moreels,~I. Near-Infrared emitting colloidal PbS
  nanoplatelets: lateral size control and optical spectroscopy. \emph{Chemistry
  of Materials} \textbf{2017}, \emph{29}, 2883--2889\relax
\mciteBstWouldAddEndPuncttrue
\mciteSetBstMidEndSepPunct{\mcitedefaultmidpunct}
{\mcitedefaultendpunct}{\mcitedefaultseppunct}\relax
\EndOfBibitem
\bibitem[Akkerman \latin{et~al.}(2019)Akkerman, Mart{\'\i}n-Garc{\'\i}a, Buha,
  Almeida, Toso, Marras, Bonaccorso, Petralanda, Infante, and
  Manna]{akkerman2019}
Akkerman,~Q.~A.; Mart{\'\i}n-Garc{\'\i}a,~B.; Buha,~J.; Almeida,~G.; Toso,~S.;
  Marras,~S.; Bonaccorso,~F.; Petralanda,~U.; Infante,~I.; Manna,~L. Ultrathin
  Orthorhombic PbS Nanosheets. \emph{Chemistry of Materials} \textbf{2019},
  \emph{31}, 8145--8153\relax
\mciteBstWouldAddEndPuncttrue
\mciteSetBstMidEndSepPunct{\mcitedefaultmidpunct}
{\mcitedefaultendpunct}{\mcitedefaultseppunct}\relax
\EndOfBibitem
\bibitem[Esser \latin{et~al.}(2000)Esser, Runge, Zimmermann, and
  Langbein]{esser2000photoluminescence}
Esser,~A.; Runge,~E.; Zimmermann,~R.; Langbein,~W. Photoluminescence and
  radiative lifetime of trions in GaAs quantum wells. \emph{Physical Review B}
  \textbf{2000}, \emph{62}, 8232\relax
\mciteBstWouldAddEndPuncttrue
\mciteSetBstMidEndSepPunct{\mcitedefaultmidpunct}
{\mcitedefaultendpunct}{\mcitedefaultseppunct}\relax
\EndOfBibitem
\bibitem[Califano \latin{et~al.}(2007)Califano, Franceschetti, and
  Zunger]{califano2007lifetime}
Califano,~M.; Franceschetti,~A.; Zunger,~A. Lifetime and polarization of the
  radiative decay of excitons, biexcitons, and trions in CdSe nanocrystal
  quantum dots. \emph{Physical Review B} \textbf{2007}, \emph{75}, 115401\relax
\mciteBstWouldAddEndPuncttrue
\mciteSetBstMidEndSepPunct{\mcitedefaultmidpunct}
{\mcitedefaultendpunct}{\mcitedefaultseppunct}\relax
\EndOfBibitem
\bibitem[Ayari \latin{et~al.}(2021)Ayari, Quick, Owschimikow, Christodoulou,
  Bertrand, Artemyev, Moreels, Woggon, Jaziri, and
  Achtstein]{ayari2021correction}
Ayari,~S.; Quick,~M.~T.; Owschimikow,~N.; Christodoulou,~S.; Bertrand,~G.~H.;
  Artemyev,~M.; Moreels,~I.; Woggon,~U.; Jaziri,~S.; Achtstein,~A.~W.
  Correction: Tuning trion binding energy and oscillator strength in a
  laterally finite 2D system: CdSe nanoplatelets as a model system for trion
  properties. \emph{Nanoscale} \textbf{2021}, \relax
\mciteBstWouldAddEndPunctfalse
\mciteSetBstMidEndSepPunct{\mcitedefaultmidpunct}
{}{\mcitedefaultseppunct}\relax
\EndOfBibitem
\bibitem[Movilla \latin{et~al.}({2020})Movilla, Planelles, and
  Climente]{MovillaJPCL}
Movilla,~J.~L.; Planelles,~J.; Climente,~J.~I. {Dielectric Confinement Enables
  Molecular Coupling in Stacked Colloidal Nanoplatelets}. \emph{{J. Phys. Chem.
  Lett.}} \textbf{{2020}}, \emph{{11}}, {3294--3300}\relax
\mciteBstWouldAddEndPuncttrue
\mciteSetBstMidEndSepPunct{\mcitedefaultmidpunct}
{\mcitedefaultendpunct}{\mcitedefaultseppunct}\relax
\EndOfBibitem
\bibitem[Lelong and Bastard(1996)Lelong, and Bastard]{lelong1996binding}
Lelong,~P.; Bastard,~G. Binding energies of excitons and charged excitons in
  GaAsGa (In) As quantum dots. \emph{Solid state communications} \textbf{1996},
  \emph{98}, 819--823\relax
\mciteBstWouldAddEndPuncttrue
\mciteSetBstMidEndSepPunct{\mcitedefaultmidpunct}
{\mcitedefaultendpunct}{\mcitedefaultseppunct}\relax
\EndOfBibitem
\bibitem[Feldmann \latin{et~al.}(1987)Feldmann, Peter, G{\"o}bel, Dawson,
  Moore, Foxon, and Elliott]{feldmann1987linewidth}
Feldmann,~J.; Peter,~G.; G{\"o}bel,~E.; Dawson,~P.; Moore,~K.; Foxon,~C.;
  Elliott,~R. Linewidth dependence of radiative exciton lifetimes in quantum
  wells. \emph{Physical review letters} \textbf{1987}, \emph{59}, 2337\relax
\mciteBstWouldAddEndPuncttrue
\mciteSetBstMidEndSepPunct{\mcitedefaultmidpunct}
{\mcitedefaultendpunct}{\mcitedefaultseppunct}\relax
\EndOfBibitem
\bibitem[Planelles \latin{et~al.}(2018)Planelles, Achtstein, Scott,
  Owschimikow, Woggon, and Climente]{planelles2018tuning}
Planelles,~J.; Achtstein,~A.~W.; Scott,~R.; Owschimikow,~N.; Woggon,~U.;
  Climente,~J.~I. Tuning intraband and interband transition rates via excitonic
  correlation in low-dimensional semiconductors. \emph{ACS Photonics}
  \textbf{2018}, \emph{5}, 3680--3688\relax
\mciteBstWouldAddEndPuncttrue
\mciteSetBstMidEndSepPunct{\mcitedefaultmidpunct}
{\mcitedefaultendpunct}{\mcitedefaultseppunct}\relax
\EndOfBibitem
\bibitem[Combescot and Shiau(2015)Combescot, and Shiau]{combescot2015excitons}
Combescot,~M.; Shiau,~S.-Y. \emph{Excitons and Cooper pairs: two composite
  bosons in many-body physics}; Oxford University Press, 2015\relax
\mciteBstWouldAddEndPuncttrue
\mciteSetBstMidEndSepPunct{\mcitedefaultmidpunct}
{\mcitedefaultendpunct}{\mcitedefaultseppunct}\relax
\EndOfBibitem
\end{mcitethebibliography}

\end{document}